\documentclass[singlecolumn]{revtex4}
\usepackage[colorlinks,bookmarksopen,bookmarksnumbered,citecolor=magenta,urlcolor=red]{hyperref}
\usepackage{graphicx}
\usepackage{amsmath}
\usepackage{natbib}
\usepackage{subfigure}
\usepackage{xcolor}
\usepackage[percent]{overpic}
\usepackage{tikz}
\usepackage{amsmath}
\usepackage{url}
\usepackage{times}
\usepackage{epstopdf}
\begin{document}
\bibliographystyle{abbrvnat}
	
\title{Influx of Bay of Bengal waters and stirring trends in the Arabian Sea based on satellite altimetry}
	
	
	
	
\author{Nihar Paul$^{1}$, Manikandan Mathur$^{2}$, Jai Sukhatme$^{3,4}$, J. Thomas Farrar$^{1}$, and Debasis Sengupta$^{5}$}

\affiliation{$^1$Woods Hole Oceanographic Institution, Woods Hole, MA, USA.}
\affiliation{$^2$Geophysical Flows Lab \& Department of Aerospace Engineering, Indian Institute of Technology Madras, Chennai, India.}
\affiliation{$^3$Centre for Atmospheric and Oceanic Sciences, Indian Institute of Science, Bengaluru, India.}
\affiliation{$^4$Divecha Centre for Climate Change, Indian Institute of Science, Bengaluru, India.}
\affiliation{$^5$International Centre for Theoretical Sciences, Bengaluru, India.}	
	
%

\begin{abstract}
\noindent 
Freshwater export from the Bay of Bengal (BoB) can drive the regional air-sea interaction in the Arabian Sea (AS). We use AVISO geostrophic and Globcurrent velocities to characterize horizontal stirring on a seasonal and interannual time scale for 1993-2022. With an example of the post-monsoon period of 2015-2016, we estimate the residence time of parcels initialized around Sri Lanka in the BoB advected to the southeastern AS is $\mathcal{O}$(1.5-2) months. Finite-time Lyapunov Exponent (FTLE) characterizes the chaotic nature of stirring through its probability density function on a sub-monthly timescale. Stirring rates are enhanced along the western boundary by 1.3 times around the Great Whirl and Socotra eddies relative to the eastern boundary and are higher in the summer monsoon season. The southeastern AS shows enhanced stirring rates during the winter monsoons. At the basin scale, the geostrophic eddy kinetic energy increases $\sim$10\% on interannual timescales associated with enhanced stirring.
\end{abstract}


\maketitle

\section*{Plain Language Summary}
\noindent We examine large-scale stirring from altimetry observation in the Arabian Sea (AS). Recent exploration suggests that AS is warming faster than the tropical oceans. We focus on the Southeastern AS during the post-monsoon seasons, when freshwater from the Bay of Bengal is transported to AS, quantifying its residence time in this region, an important metric to understand the regional air-sea interaction and monsoon onset. Concomitantly, using Lagrangian measures, we diagnose stirring from altimetry observations, such as the surface geostrophic and Globcurrent datasets, comprising thirty years. We statistically characterize horizontal stirring between the western and eastern AS and the entire basin and showcase an enhancement in the long-term trend. The study provides an improved understanding of the lateral processes in the AS from seasonal to inter-annual scales and brings forth a reliable method to evaluate the performance of the ocean general circulation models in quantifying stirring.

\section{Introduction}

\noindent Advective transport and stirring exert a profound influence on a wide range of upper ocean processes \cite{koshel2006chaotic,von1999lateral,naveira2011eddy,fernandez2024isopycnal}, including bio-oceanography \cite{abraham1998generation,martin2003phytoplankton}. Examples of such bio-oceanography phenomena also include the formation of highly concentrated green buoyant cyanobacteria phytoplankton blooms in the Baltic Sea \cite{d2018ocean} and chlorophyll-a patchiness in the Atlantic Ocean \cite{mahadevan2002biogeochemical,lehahn2007stirring}. 
\citet{schmidt2020seasonal} outlined the seasonal changes in the advective pathways of the oxygen minimum zone in the eastern basin of AS. In the North Indian Ocean (NIO), advective transport and stirring in the Arabian Sea (AS), including the cross-basin exchange with the Bay of Bengal (BoB), are recognized to affect the regional weather and climate \cite{schott2001monsoon,schott2009indian,beal2013response,phillips2021progress}. In this work, we used decades-long satellite altimetry to objectively characterize (i) transport of BoB freshwater into the AS and (ii) basin-scale stirring of passive tracers in the AS.\\

\noindent Before the onset of the South Asian summer monsoon in June, most of the AS is a part of the Indian Ocean warm pool, the warmest region in the world's oceans at this time. In particular, the southeastern AS warms in late April-May, with sea-surface temperature (SST) of 28-30$^\circ$C \cite{vinayachandran2007arabian,neema2012characteristics}, and supports large-scale, organized atmospheric convection at the time of monsoon onset \cite{gadgil1984ocean,waliser1993convective}. After the onset, the southwesterly seasonal wind pattern brings heavy rainfall to the Indian subcontinent, and the western AS cools rapidly due to upwelling and enhanced evaporation \cite{vinayachandran2007arabian,izumo2008role}. On longer time scales, the AS has recently been warming at 0.3$^\circ$C per decade (significantly faster than the global tropics), with distinct multidecadal modulation \cite{sun2019recent}. Observations also indicate a corresponding increase in intense tropical cyclones \cite{deshpande2021changing,maneesha2023intrusion} and marine heat waves \cite{saranya2022genesis,chatterjee2022marine,koul2023seasonal}.\\

\noindent The southeastern AS is also characterized by the presence of an anti-cyclonic eddy with ``high'' sea-level anomaly that forms during January (mid-northeast monsoon) to February called ``Lakshadweep High'' along with the Western Indian Coastal Current (WICC) and Kelvin waves \cite{shankar1997dynamics}. It is known to affect the high sea-surface temperature before monsoon onset \cite{shenoi1999sea}. The influx of freshwater from the Bay of Bengal (BoB) to the AS during the northeast monsoon modifies the vertical stratification of the southeastern AS, affecting the regional air-sea interactions \cite{schott1994currents,mathew2018dynamics}. This leads to the formation of a thick barrier layer with an inversion in temperature \cite{shankar2004observational,thadathil2008seasonal,echols2020spice} and can support the warming of the southeastern AS before the monsoon onset \cite{shenoi2004remote,durand2004impact}. Modeling studies, however, suggest that the lateral advection of low saline water during winter monsoon has a weak effect on warming of the southeastern AS before the monsoon onset \cite{vinayachandran2007arabian,akhil2023southeastern}. In addition, coupled models do not capture the transport of low-salinity water from the BoB, resulting in a mixed layer depth (MLD) and thermocline depth bias in the southeastern AS \cite{nagura2018origins}.\\

\noindent The seasonal pathways of near-surface freshwater from Bay of Bengal (BoB) runoff and Indonesian Throughflow (ITF) to different regions of the Indian Ocean using climatological observations have been studied by \citet{sengupta2006surface}. In recent two decades, sea surface salinity in the southeastern AS declined due to increased transport of low-salinity water with intensified West India coastal currents during winter \cite{varna2021strengthening}. While the aforementioned studies highlight prominent features of the transport of BoB waters to the southeastern Arabian Sea, a systematic Lagrangian analysis is required to objectively/rigorously characterize the transport. Indeed, such a Lagrangian approach has revealed the seasonal to interannual variability in advective pathways and transit times of particles in the Red Sea outflow to different regions of AS \cite{menezes2021advective,menezes2023interannual}. In this study, we build on a similar approach to study the residence time of parcels when they leave the BoB and enter the southeastern AS region during the winter monsoon. Furthermore, we identify kinematic barriers in the flow of the southeastern AS using the Lagrangian descriptor M-function metric \cite{mancho2013lagrangian,agaoglou2024building} and to examine the fate of parcels when released from BoB. During their residence in the region, passive tracers can be stirred and trapped by mesoscale eddies, likely as in north BoB \cite{paul2021eddy,paul2023eddy}, boundary currents and waves and their pathways are characterized using the M-function. We particularly present a case for 2015-2016 showing the evolution of the residence time of the parcels on the monthly timescale released from BoB from October to March in the southeastern AS.\\

\noindent Lagrangian tools from dynamical systems such as Finite-time Lyapunov Exponent (FTLE) have proved to be useful in quantifying stirring through simple advective protocols in multiscale turbulent flows \cite{mathur2007uncovering,haller2015lagrangian,aref2020stirring}. Moreover, stirring can introduce fine scales and drive mixing \cite{eckart1948analysis,ottino1989kinematics,muller2002stirring}. Further, Finite-size Lyapunov Exponents were also helpful in interpreting chlorophyll patterns in the western Arabian Sea \cite{kumar2023study}. Therefore, the question is how important are the advective effects on FTLE metric evolving over the seasons, their spatial distribution, and the probability density function \cite{sukhatme2002decay, waugh2008stirring,paul2020seasonality} in the AS basin? Given the trend in warming, does it affect the stirring rates on the basin scale at the inter-annual timescale?\\

\noindent The paper is organized as follows, with section \ref{sec:2} presenting the data and methods. In section \ref{sec:3.1}, we estimate the upper ocean volume transport between BoB and AS across the channel within the MLD. We present a case study from 2015 to 2016 to illustrate the Lagrangian pathway of tracers released around Sri Lanka to the different regions of the NIO during the northeast monsoon season. We estimate the residence timescale of parcels in the southeastern AS and showcase the kinematic barriers in the flow using the M-function descriptor. We use Finite-time Lyapunov Exponents in section \ref{sec:3.2} to show the associated statistics with basin-scale trends for enhanced eddy kinetic energy and stirring by surface geostrophic flow. The manuscript ends with a summary of the results in section \ref{sec:4}.\\

\begin{figure}[hbt]
\centering
\includegraphics[width=\textwidth]{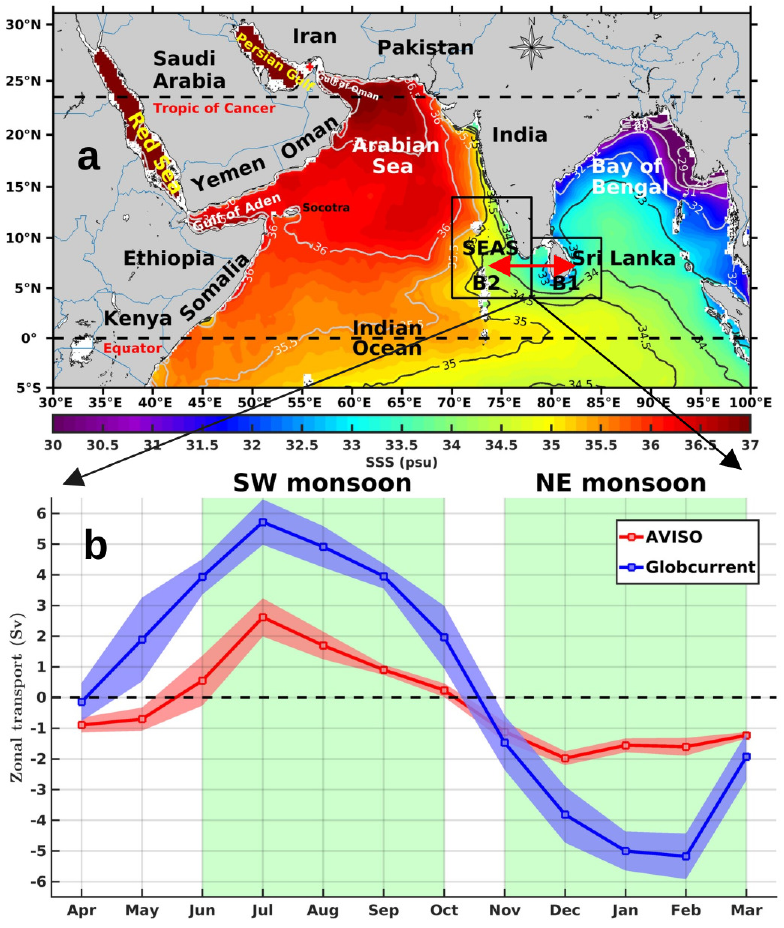}
\caption{(a) Soil Moisture Active Passive (SMAP) sea-surface salinity 2015-2022 climatology for the north Indian Ocean (30$^\circ$E-100$^\circ$E, $5^\circ$S-31$^\circ$N). The region surrounding Sri Lanka is marked by box B1 (78$^\circ$E-85$^\circ$E, 4$^\circ$N-10$^\circ$N), and Southeastern Arabian Sea (70$^\circ$E-78$^\circ$E, 4$^\circ$N-14$^\circ$N) is enclosed by box B2. (b) Average zonal transport by surface geostrophic (red) and Globcurrent (blue) velocities in Sv (10$^6$ m$^3$ s$^{-1}$) integrated over the mixed layer depth across the boundary at 78.125$^\circ$E, 4.125$^\circ$N-8.625$^\circ$N (shown from April to May from 2015 to 2022). The shaded region in red and blue denotes the standard deviations, respectively. The red arrow in (a) shows the eastward and westward directions of transport across the boundary of the boxes during southwest (SW) and northeast (NE) monsoons.}
\label{fig1}
\end{figure}

\section{Data sources and methods}\label{sec:2}

\noindent We use the daily AVISO surface geostrophic, Globcurrent velocities, and sea-level anomaly datasets gridded at 0.25$^\circ$ to characterize stirring during 1993-2022. The geostrophic currents along the Equator are estimated using the $\beta$-plane approximation \cite{moore1977modeling,lagerloef1999tropical}. The coarse resolution of altimetry-based surface currents does not allow assessment of scales less than 100 km \cite{arbic2013eddy,dufau2016mesoscale}. The surface Globcurrent velocity is the sum of surface geostrophic current and an Ekman component derived using European Centre for Medium-Range Weather Forecasts Reanalysis (ERA) wind forcing and measurements from surface drifters and Argo ﬂoats \cite{rio2014beyond}. A comparison of the Globcurrent velocities with the Doppler Volume Sampler Acoustic Doppler Current Profile current at 1.25 m of the mooring-AD10 (SI Figure. S1) gives a regression coefficient of $\sim 0.8$. 8-day running mean sea surface salinity (SSS) from the Soil Moisture Active Passive (SMAP) satellite \cite{fore2016combined}, and sea-surface temperature (SST) from Optimum Interpolation Surface Temperature (OISST) have been used from 2015 to 2022 for the analysis. Density has been computed using the Gibbs-Sea-Water Oceanographic Toolbox \cite{mcdougall2011getting}. Argo MLD monthly climatology binned to 1-degree \cite{holte2017argo} has also been used to estimate depth-integrated transport from the altimetry dataset. The evolution of the position of the tracers is given by $d\lambda/dt=u(\lambda,\phi,t)/Rcos(\phi)$, and $d\phi/dt=v(\lambda,\phi,t)/R$, where $\lambda$ and $\phi$ are the longitude and latitudes of the particles, and R is the earth's radius (6371 km). The 4th ordered Runge Kutta scheme has been used to advect Lagrangian particles by the zonal and meridional velocities (u,v) on a fixed grid in time (t) and a cubic interpolation scheme to interpolate the currents in space.\\



\begin{figure}[hbt]
	\centering
	\includegraphics[width=\textwidth]{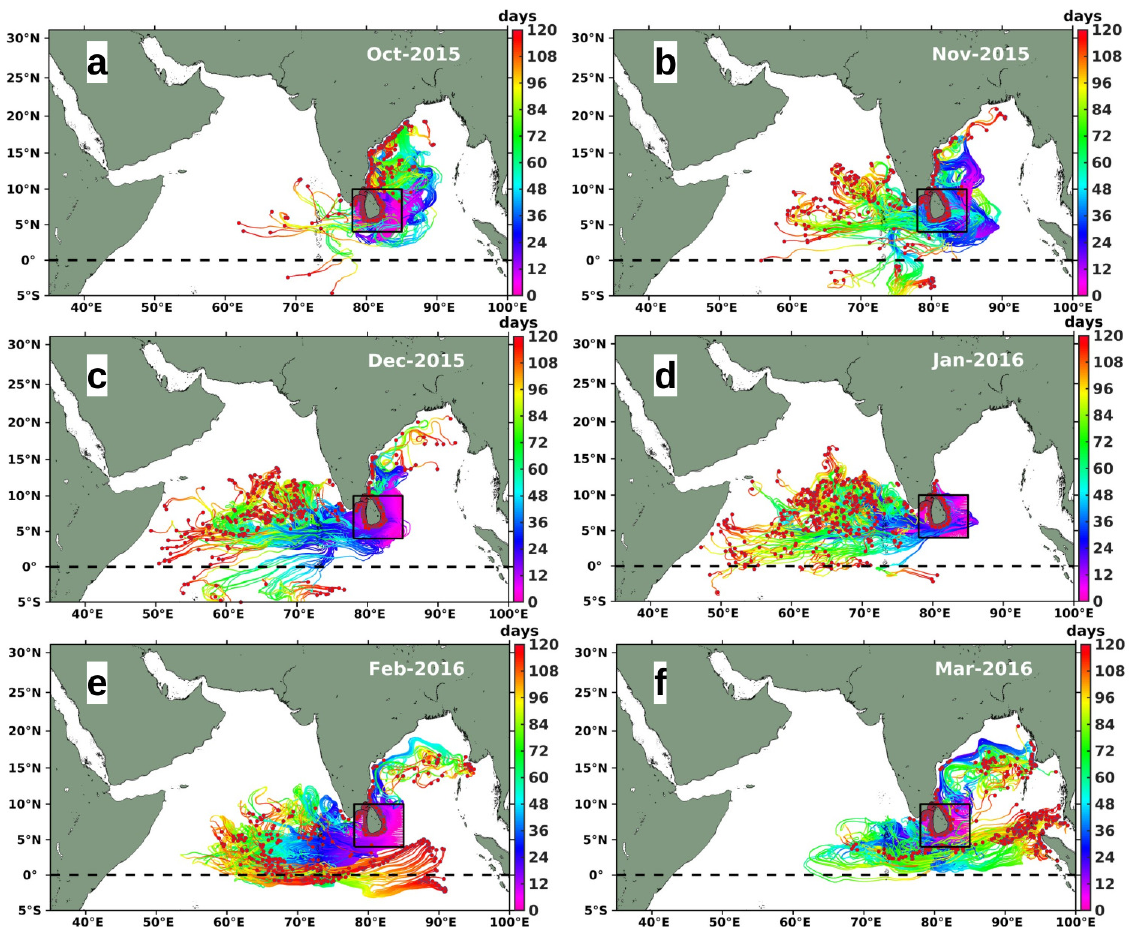}
	\caption{(a)-(f) Trajectories of passive tracers released on October 1, 2015, to March 1, 2016, initialized at a resolution of $0.25^\circ$ in box B1 advected by surface Globcurrent velocities for 120 days (colormap showing the time from release). The red dots are the final positions after 120 days of advection. The black dashed line marks the Equator.}
	\label{fig2}
\end{figure}

\section{Transport of passive tracers from the BoB to AS}\label{sec:3}

\noindent Figure \ref{fig1}(a) shows the winter period (November to February) climatological (2015-2022) map of sea surface salinity in the NIO, highlighting the remarkable contrast in freshwater content between the AS and the BoB. The boxes B1 (right) and B2 (left) were chosen based on the regional importance of water mass exchange between BoB and AS and air-sea interactions affecting the Indian monsoon onset in the southeastern AS \cite{durand2004impact,shenoi2005hydrography,kurian2007mechanisms,wu2012air,roman2020role}. The monthly climatology (2015-2022) of zonal volume transport, defined as $M_x=\int_{-MLD}^{0}u(x,y,z,t)\delta z \delta y$ integrated over the MLD assuming equivalent barotropic flow across the boxes (the boundaries mentioned in Figure \ref{fig1} caption) from surface geostrophic and Globcurrent velocities is shown in Figure \ref{fig1}(b). The monthly climatology of the MLD across the NIO and the channel within the boxes are shown in SI Figure S2, S3. During the SW and NE monsoon, the upper ocean transport within the MLD by surface geostrophic and Globcurrent velocities is eastward and westward, respectively, with the maximum values of around 2.6 Sv and 5.7 Sv.\\


\noindent We release virtual tracers from box B1 to show the Lagrangian pathways to the different regions of NIO from October 2015 to March 2016 (Figure \ref{fig2}). The tracers are initialized at a resolution of $0.25^\circ$ and advected for 120 days following the intra-seasonal period \cite{cheng2013intraseasonal} starting on day 1 of each month. During October, the final position of the particles (marked in red dots with the trajectory) is aligned along the western coast of BoB, with a few of them entering the southeastern AS region while the remaining cross the Equator. With the change in the seasonal flow, the particles start moving towards the southeastern AS from November to January; the trajectories do not reach the western coast of BoB during January. Moreover, a significant fraction of tracers are transported to the region closer to Somalia within 120 days during December and January.
In contrast to earlier months, during February and March, the tracers exhibit minimal dispersion while moving eastward along the equatorial wave-guide towards the Sumatra islands. The return flow also enhances the transport of particles to the interior of BoB in February and March. A similar numerical experiment was performed from April 2016 to September 2016, shown in SI Figure S4, and the result suggests tracers are transported to the interior of the BoB except during September 2016 when the parcels started moving westward southwest of Sri Lanka.\\ 

\begin{figure}[hbt]
	\centering
	\includegraphics[width=\textwidth]{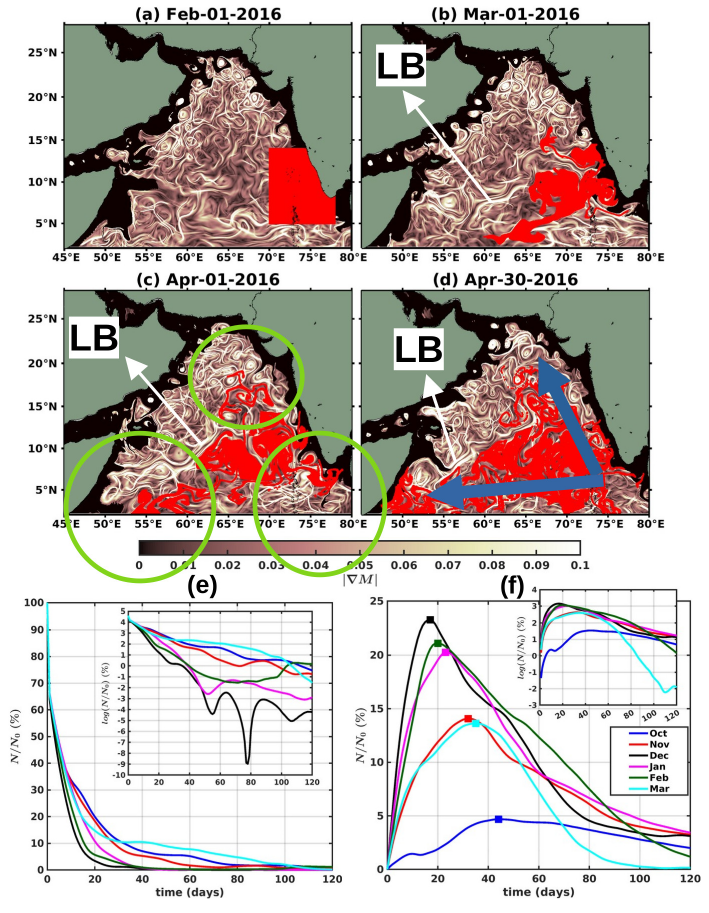}
	\caption{(a)-(d) shows the evolution of parcels initialized on box B2 (resolution of 1 km) shown on February 1, March 1, April 1, and April 30 of 2016 in the Southeastern Arabian Sea overlaid on M-function with advection time (forward and backward) of two weeks (computational grid of $\frac{1}{55}^\circ$). The white arrow shows the Lagrangian barrier abbreviated as ``LB'' in (b), (c), (d), and the green circle showcases the region that experiences filamentation in (c). The blue arrow in (d) denotes the overall direction of dispersion. (e), and (f) represents the monthly number density of parcels (\%) in boxes B1 and B2 initialized on box B1 shown at a resolution of 1 km on all the days from October to March advected for 120 days by Globcurrent velocities in 2015-2016 (the y-axis of the insets are shown in logarithmic scale). A square marker symbolizes the maxima of the curves in (f).}
	\label{fig3}
\end{figure}

\subsection{M-function and barrier to transport}\label{sec:3.1}

\noindent We now identify kinematic barriers in the Globcurrent flow using the Lagrangian descriptor M-function; the tool has been used to model Rossby wave breaking in the Southern Spring Stratosphere \cite{guha2016modeling}, understanding the mixing of warmer and colder air-parcels during lifting (updraft \& downdraft, analogous to upwelling, and downwelling in the ocean) in the tropical storm \cite{rutherford2012lagrangian,niang2020transport}, and likely be helpful to identify a barrier to transport in the AS. The M-function is defined as follows:

\begin{equation}
	M(\mathbf{x_0},t_0,\tau)=\int_{t_0-\tau}^{t_0+\tau}||\mathbf{v}(\mathbf{x}(t;\mathbf{x_0}),t))||dt.\label{eqn:11}
\end{equation}

\noindent where $\mathbf{v(x,}\: t)$ is the velocity field and $||\cdot{}||$  denotes the Euclidean norm. At a given time $t_0, M$ corresponds to the length of the trajectory traced by a fluid parcel starting at $\mathbf{x_0} = \mathbf{x}(t_0)$ as it evolves forward and backward in time for a time interval $\tau$. For sufficiently large $\tau$ values, the sharp changes
that occur in narrow gaps in the scalar field provided by M, which we will refer to as singular features, highlight the stable and unstable manifolds and hyperbolic trajectories at their crossings. The computational grid of $\frac{1}{55}^\circ$, and $\tau$ as two weeks are taken for the M-function estimates. We show the evolution of passive tracers initialized on box B2 at a resolution of 1 km initialized on February 1, 2016, advected for 90 days in Figure \ref{fig3}(a)-(d). The parcels are stretched and folded, forming filament as the advecting Globcurrent flow transports them (shown in the green circle in Figure \ref{fig3}(c)). Some parcels remain as a blob of the size of the mesoscale eddies in the central AS and are then transported to the interior, thus having different advecting timescales. We also notice the presence of the Lagrangian barrier (LB; marked in white arrow in Figure \ref{fig3}(b)-(d)), inhibiting the exchange of tracers across the manifolds while the parcels are transported to the southern-western AS. Overall, the pictures suggest myriads of dynamics in the surface flow of the AS while the tracers move northward and westward, as indicated by the arrow in Figure \ref{fig3}(d). \\ 

\noindent We then evaluate the fate of parcels released from box B1 and quantify their residence time in southeastern AS box B2 during the winter monsoon of 2015-2016. The parcels are initialized on a grid resolution of 1 km in box B1 and released on all days from October to March, advected for 120 days. The fraction of particles denoted by $N/N_0$ in $\%$ averaged over the month for boxes B1 and B2 are shown in Figure \ref{fig3}(e),(f). The number of parcels in box B1 decays approximately exponentially with an e-folding time of 8, 8, 5, 8, 6, and 7 days from October to March. A significant fraction of the parcels enter from box B1 to B2, and the fraction becomes maximum and decays over time. The number of parcels in box B2 is given by $N(t)=N_0exp(-(t-\tau_{max})/\tau_r)$, where $\tau_{max}$, $\tau_r$ is the time to reach maxima, and the residence time based on the e-folding period, and $N_0$ is the maximum number of parcels at $\tau_{max}$ \cite{pentek1996transient}. The maxima for the curves occur over $\tau_{max}$$\sim$ 45, 33, 18, 24, 21, and 36 days, and the corresponding residence time $\tau_{r}$$\sim$ $>76$, 53, 46, 50, 58, and 31 days from October to March. The exchange is $5\%$ for October, 10-15$\%$ for November, and March, 20-25$\%$ for December and January, respectively. The number suggests the exchange is maximum from December to January, and the residence time of the parcels is 1.5 to 2 months in the southeastern AS.\\

\subsection{Finite-time Lyapunov Exponents and the basin-scale trends in the Arabian Sea}\label{sec:3.2}

\noindent In a time-dependent chaotic flow, we use the Finite-time Lyapunov Exponent \cite{benettin1980lyapunov} to characterize stirring. It is defined as the exponential rate of separation from an initial infinitesimal separation when advected over a finite period and is represented by Equation \ref{eqn:3} reads as,\\

\begin{equation}
	\lambda_\tau(\boldsymbol{x_0})=\frac{1}{\tau}log(\frac{||\boldsymbol{\delta x}(t)||}{||\boldsymbol{\delta x}(t_0)||}).\label{eqn:3}
\end{equation}

\noindent where, $\boldsymbol{\delta x}(t_0)$, $\boldsymbol{\delta x}(t)$, are the separation at time $t_0$, and $t$, and $\tau$ is time of advection. $F^t_{t_0}(\boldsymbol{x_0})$ denotes the position of a parcel at time $t$ forward in time, advected by the flow from an initial time and position ($t_0,\boldsymbol{x_0}$). $C^{t}_{t_0}(\boldsymbol{x_0})$ is the Cauchy green Lagrangian tensor, which is symmetric and positive definite where the eigenvalues ($\lambda's$) and eigenvectors ($\xi's$) can be written as,\\

\begin{align}
	&C^{t}_{t_0}(\boldsymbol{x_0}) = {(\nabla F^t_{t_0}(\boldsymbol{x_0}))}^T \nabla {F^t_{t_0}(\boldsymbol{x_0})},\label{eqn:4}\\
	&C^{t}_{t_0}(\boldsymbol{x_0})=\lambda_i\xi_i,\: 0<\lambda_1\le\lambda_2, i= 1, 2.\label{eqn:5}
\end{align}

\noindent The gradient of the flow map $\nabla F^t_{t_0}(\boldsymbol{x_0})$ is calculated about an auxiliary reference relative to the computational grid points \cite{onu2015lcs} and can be written as \\

\begin{align}
	\nabla F^t_{t_0}(\boldsymbol{x_0})\approx \begin{pmatrix} 
		\alpha_{11} & \alpha_{12} \\
		\alpha_{21}& \alpha_{22}
	\end{pmatrix},
	\label{eqn:6}
\end{align}
where, 

\begin{align}
	\alpha_{i,j}\equiv \frac{x_{i}(t;t_0,x_0+\delta x_j)-x_{i}(t;t_0,x_0-\delta x_j)}{2|\delta x_j|}.
	\label{eqn:7}
\end{align}

\noindent Therefore, the largest FTLE in forward direction also called f-FTLE \cite{haller2002lagrangian,haller2011lagrangian,mathur2019thermal} associated with the trajectory $\boldsymbol{x}(t,t_0,\boldsymbol{x_0})$ over the time interval [$t_0$, $t$] is given by,\\

\begin{align}
	\lambda_{\tau}(\boldsymbol{x_0})=\frac{1}{|t-t_0|}\log(\sqrt{\lambda_{\max}[C^{t}_{t_0}(\boldsymbol{x_0})]}).
	\label{eqn:8}
\end{align}

\begin{figure}[hbt]
	\centering
	\includegraphics[width=\textwidth]{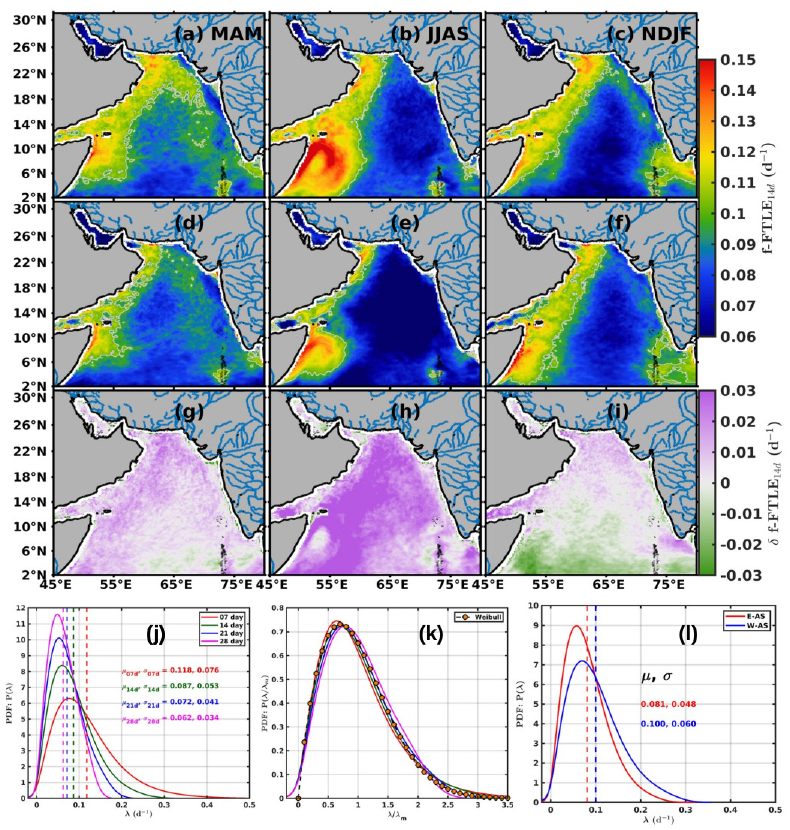}
	\caption{(a)-(c), (d)-(f) represents the seasonal climatology for March-April-May (MAM; summer), June-July-August-September (JJAS; southwest monsoon), and November-December-January-February (NDJF; north-east monsoon) of Finite-Time Lyapunov Exponent with an integration time of 2 weeks using surface geostrophic and Globcurrent current, respectively. (g)-(i) shows the difference in rows one and two estimates for the corresponding seasons. (j) shows Probability Density Function (PDF) of Lyapunov Exponent ($\lambda$, f-FTLE, units in day$^{-1}$) for an integration time of a week to 4 weeks estimated from the surface geostrophic current (record length beginning from 2013-2022) for the Arabian Sea; the mean ($\mu$) and the standard deviation ($\sigma$) in day$^{-1}$ are mentioned in the text. (k) shows the normalized PDF (by the mean) for all the integration times, and curves are fitted with a Weibull distribution with parameters a=1.1 and b=1.8. (l) shows the PDF of f-FTLE for the eastern and western Arabian Sea (E-AS; 60.125$^\circ$E-80.125$^\circ$E, 2$^\circ$N-31.125$^\circ$N and W-AS; 45$^\circ$E-60$^\circ$E, 2$^\circ$N-31.125$^\circ$N) for the integration time of 14 days with their mean and standard deviation.}
	\label{fig4}
\end{figure}

\noindent The forward integration time $|\tau|=|t-t_0|$ has been taken as 7, 14, 21, and 28 days and computed on a finer grid resolution of $\frac{1}{8}^\circ\times\frac{1}{8}^\circ$ on the domain 45$^\circ$E-80$^\circ$E, 2$^\circ$N-31$^\circ$N. Using an example of an anti-cyclonic eddy (Lakshadeep High) in the southeastern AS on February 1, 2016, we showcase the difference between FTLEs with backward (b-FTLE or a-LCS; attracting Lagrangian Coherent Structure) and forward (f-FTLE or r-LCS; repelling LCS) integration time of 2 weeks and the corresponding M-function and its gradient to highlight the kinematic barrier in surface flow of Globcurrent in SI Figure S5. The March-April-May (MAM), June-July-August-September (JJAS), and November-December-January-February (NDJF) climatology estimates for the f-FTLE with an advection time of 14 days are shown in Figure \ref{fig4}. Panel (a)-(c), (d)-(f) show the estimates from surface geostrophic current and Globcurrent flow field, and panel (g)-(i) shows the differences in the seasons. The chaotic nature of horizontal stirring shows enhancement in the western boundary compared to the eastern boundary of the AS in all seasons. The highest magnitude of f-FFTLE-14 days is in the region containing the Great Whirl and Socotra eddy closer to Somalia in the monsoon season. A seasonal cycle persists with increasing f-FTLE from MAM to JJAS and gradually decreasing during NDJF on the western boundary. Unlike the eastern boundary, the f-FTLE in the northeastern AS is high during MAM and remains quiescent during the other seasons, except during the winter monsoon period, when the values become high near the southern tip of India around the southeastern AS. Further, the central AS shows relatively low values of f-FTLEs due to the dominance of vorticity over the strain in the flow (not shown). These estimates are quite robust in both the surface geostrophic and Globcurrent velocities. The difference in the estimates of the f-FTLE suggests that geostrophic velocity stirs more than the Globcurrent velocity field horizontally in all the seasons except for the southern AS during the winter monsoon.\\    

\noindent We now characterize the probability density function (PDF) of f-FTLE from 2013-2022 for the AS basin. The PDF has a stretched exponential character and a long tail, which becomes fatter with integration time, ranging from a week to four weeks. The mean and standard deviation of the f-FTLE decreases with an increase in integration time shown in Figure \ref{fig4}(j). For consistency with other parts of the world's oceans, such as the Tasman Sea and the BoB \cite{waugh2008stirring,waugh2012diagnosing,paul2020seasonality}, we note that a Weibull distribution approximately fits the FTLE histogram normalized by the mean FTLE for the year 2013-2022. The specific expression plotted in Figure \ref{fig4}(k) satisfies,\\

\begin{equation}
	P_W(\lambda) = \frac{b}{a}(\frac{\lambda}{a})^{b-1} exp(-\frac{\lambda^b}{a^b}).\label{eqn:9}
\end{equation} with a = 1.1 and b = 1.8.\\

\noindent We further characterize the western and eastern AS (W-AS and E-AS) defined by region by 45$^\circ$E-60$^\circ$E, 2$^\circ$N-31.125$^\circ$N, 60.125$^\circ$E-80.125$^\circ$E, 2$^\circ$N-31.125$^\circ$N, respectively, by comparing the PDF of f-FTLE for integration time of 2 weeks in Figure \ref{fig4}(l). The PDF suggests that the f-FTLE in W-AS for all integration times (SI Figure S6) has a longer tail than the E-AS and a less stretched exponential character. Indeed, the tail becomes heavier/fatter with the increase in the integration time and the decrease in mean and standard deviation. The estimate also suggests that the W-AS is 1.3 times more chaotic than E-AS for an integration time of two weeks by taking the ratio of their mean. Similar PDFs for the basin and regional scales were constructed from 1993 to 2002 and 2003 to 2012 SI Figures S7 and S8 and showcase complementary characteristics.\\

\begin{figure}[hbt]
	\centering
	\includegraphics[width=\textwidth]{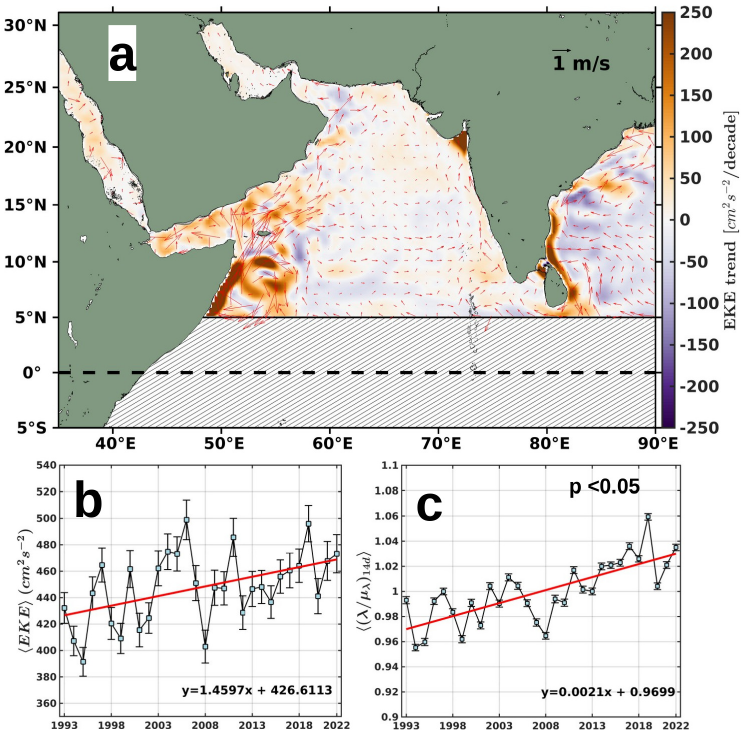}
	\caption{(a) Spatial map of the trend of daily geostrophic Eddy Kinetic Energy (EKE) overlaid with the mean current for the record length from 1993-2022 for the domain 35$^\circ$E-90$^\circ$E, 5$^\circ$S-31.125$^\circ$N. (b), and (c) shows the yearly time series of mean eddy kinetic energy and normalized f-FTLEs (by the mean for the entire record length) for an integration time of two weeks for the Arabian Sea (45$^\circ$E-80$^\circ$E, 5$^\circ$N-31.125$^\circ$N) with 95\% confidence intervals. The slope and the y-intercept of the straight line ($y=mx+c$) characterize the linear trend (in red) shown in the text of Figures (b) and (c). The statistical significance of the trend has been tested following Mann Kendall test of the null hypothesis with $\alpha=0.05$. In all the cases, the p-values are less than 0.05, indicating a monotonic increase in the time series trend. The horizontal dashed line in (a) marks the Equator, and the region covering from 5$^\circ$S to $5^\circ$N is masked.}
	\label{fig5}
\end{figure}

\noindent We then compute the Eddy Kinetic Energy (EKE), defined as,  

\begin{align}
	EKE=\frac{1}{2}(u^{\prime 2}+v^{\prime 2}).\label{eqn:10}
\end{align}

\noindent where, $u^\prime$, $v^\prime$ is the deviation of daily velocities from the long-term mean of thirty years of geostrophic currents. The 30-year-long spatial map of trend in the EKE is shown in Figure \ref{fig5}(a), suggesting an enhancement in the western boundary of the AS in the Great Whirl, Socotra eddy regions, and the Gulf of Aden. The Southeastern and northern AS show weak trends except in the Gulf of Khambhat compared to the west, with an overall increase in the basin scale yearly mean EKE shown in Figure \ref{fig5}(b). We further examined the time series of the normalized yearly mean f-FTLEs (by the long-term mean) for an integration time of 2 weeks for the entire period in Figure \ref{fig5}(c), and at the basin scale it shows an enhancement in the long-term trend at per with EKE. The integration times of a week, three weeks, and four weeks for f-FTLEs show similar agreement (SI Figure S9(a),(b), and (c)). The stirring rate shows an enhancement of $\sim5\%$ with a positive linear trend and interannual variability.\\

\section{Conclusion}\label{sec:4}

\noindent We present a comprehensive analysis of the surface stirring of the AS using AVISO and Globcurrent datasets spanning thirty years. The mean saltiness for AS is greater than 35 psu, and at least three psu is more than BoB. In particular, we chose box B1, enclosing Sri Lanka based on the seasonal climatology of sea-surface salinity, and the B2, representing southeastern AS, which is dynamically crucial in setting the onset of the Indian summer monsoon. During the SW and NE monsoon, the upper ocean transport within the MLD by surface geostrophic and Globcurrent velocities is eastward and westward, with maxima of the $\mathcal{O}(2.6)$ Sv, and $\mathcal{O}(5.7)$ Sv, respectively. We then showcase the Lagrangian pathways of particles advected with the Globcurrent flow velocities released from box B1 to different regions of the Indian Ocean from the 1st day of October 2015 to March 2016. The pathway suggests enhanced inflow to the AS from November to February. The path bifurcates when parcels are released in February and March towards the Sumatra islands and to the interior of BoB. The M-function descriptor distinctly showcases the barrier in the flow, inhibiting transporting tracers across them in the Southwestern AS when parcels are released from box B2 on February 1 and are advected for 90 days. To quantify the inflow from BoB to AS, we calculate the fraction of particles transported from box B1 to B2 during the winter monsoon of 2015-2016. The result suggests that transport is maximum from December to January, and the residence time of the parcels is 1.5 to 2 months with leading transport of particles of 20-25$\%$ to B2 in the southeastern AS.\\

\noindent We then characterize the stirring using Lagrangian metrics Finite-Time Lyapunov Exponents to quantify the chaotic nature of the flow driven by surface geostrophic and Globcurrent velocities. Both the geostrophic and Globcurrent FTLEs with two weeks of advection suggest enhanced stirring along the western boundary of the AS in contrast to the eastern boundary, highest during the JJAS around the Great Whirl and Socotra eddies. The southeastern AS lits up during the northeast monsoon. The central AS showcase is relatively weakly chaotic to the other regions, and there is a clear distinction between the western and eastern boundaries of the basin.
The probability density function of FTLEs quantitatively suggests the western AS is approximately 1.3 times more chaotic than the eastern AS. On a basin scale, the surface geostrophic FTLEs advected for a week to four weeks, suggesting it increases on an interannual timescale. It showcases enhancement with the increased eddy-kinetic energy, particularly in the western boundary around the Great Whirl and Socotra eddies. Eddy kinetic energy for the geostrophic flow is enhanced $\sim$10\% over three decades at the basin scale with increased stirring rates. These estimates can provide a reference to represent stirring for the ocean general circulation model with high fidelity.\\


\bibliography{reference_nihar.bib}

\section*{Open Research}

\noindent The AVISO surface geostrophic, Globcurrent velocities at 0 $m$, and sea-level anomaly datasets can be obtained from \url{https://doi.org/10.48670/moi-00148}, and \url{https://doi.org/10.48670/mds-00327}. The NASA Jet Propulsion Laboratory Soil Moisture Active Passive sea-surface salinity Version 5.0 and Optimum Interpolation sea-surface temperature (SST) Version 2.1 datasets are available from \url{https://doi.org/10.5067/SMP50-2TOCS}, and \url{https://doi: 10.1175/JCLI-D-20-0166.1}. The INCOIS and NIOT mooring data (AD10) are available at \url{https://incois.gov.in/portal/datainfo/mb.jsp}. The Argo mixed layer climatology is available at \url{https://10.1002/2017GL073426}. The trends associated with eddy kinetic energy and the FTLE metric have been computed using the Climate Data toolbox \cite{greene2019climate}.

\section*{Acknowledgments}
\noindent NP thanks Dr. Michael Spall and Dr. Viviane V Menezes for their helpful discussions and suggestions. JS and NP thank the Divecha Centre for Climate Change and the Indian Institute of Science, Bengaluru, India, for their financial support. NP and MM acknowledge the financial support from the grant (no. SB20210806AEMHRD008563) and project (no. SP22231241CPETWOGFLHOC) from the Ministry of Earth Sciences and Geophysical Flows Lab, India Institute of Technology Madras, Chennai, India. JS and DS recognize the University Grants Commission (UGC) for funding via 6–3/2018 under the 4th cycle of the Indo-Israel joint research program and the National Monsoon Mission, Indian Institute of Tropical Meteorology, Pune, India. The authors declare no conflict of interest and that the work has not been published elsewhere.

\section*{Supplementary Information}

\begin{enumerate}
	\item Figures S1 to S9.
\end{enumerate}
\setcounter{figure}{0}
\renewcommand{\figurename}{Fig.}
\renewcommand{\thefigure}{S\arabic{figure}}
\begin{figure}
	\centering 
	\includegraphics[scale=1]{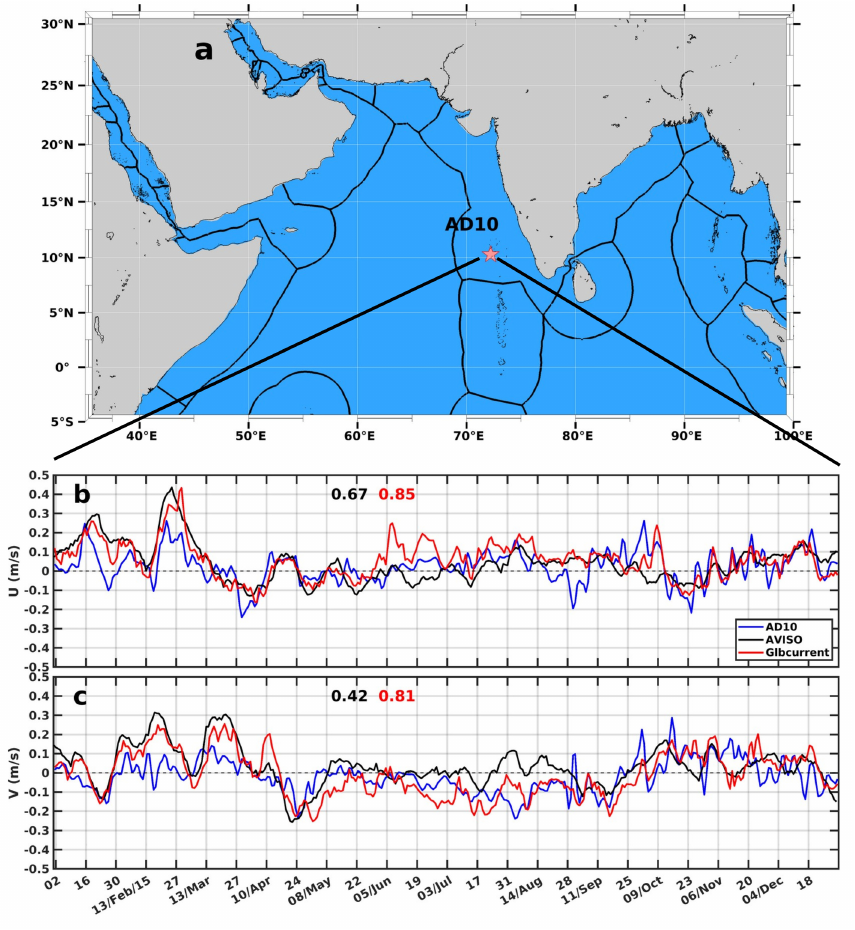}
	\caption{(a) Position of the mooring AD10 at 72.2$^\circ$E, 10.3$^\circ$N; (b), and (c) shows the comparison of the surface geostrophic and Globcurrent with the current at the mooring site from Doppler Volume Sampler (DVS) Acoustic Doppler Current Profile (ADCP) at a near-surface depth of 1.25 m. The regression slope of the surface geostrophic and Globcurrent with the near-surface DVS mooring currents (zonal and meridional) are mentioned in the text (black and red) in the subpanels (b) and (c).}
	\label{S1}
\end{figure}

\begin{figure}
	\includegraphics[width=\textwidth]{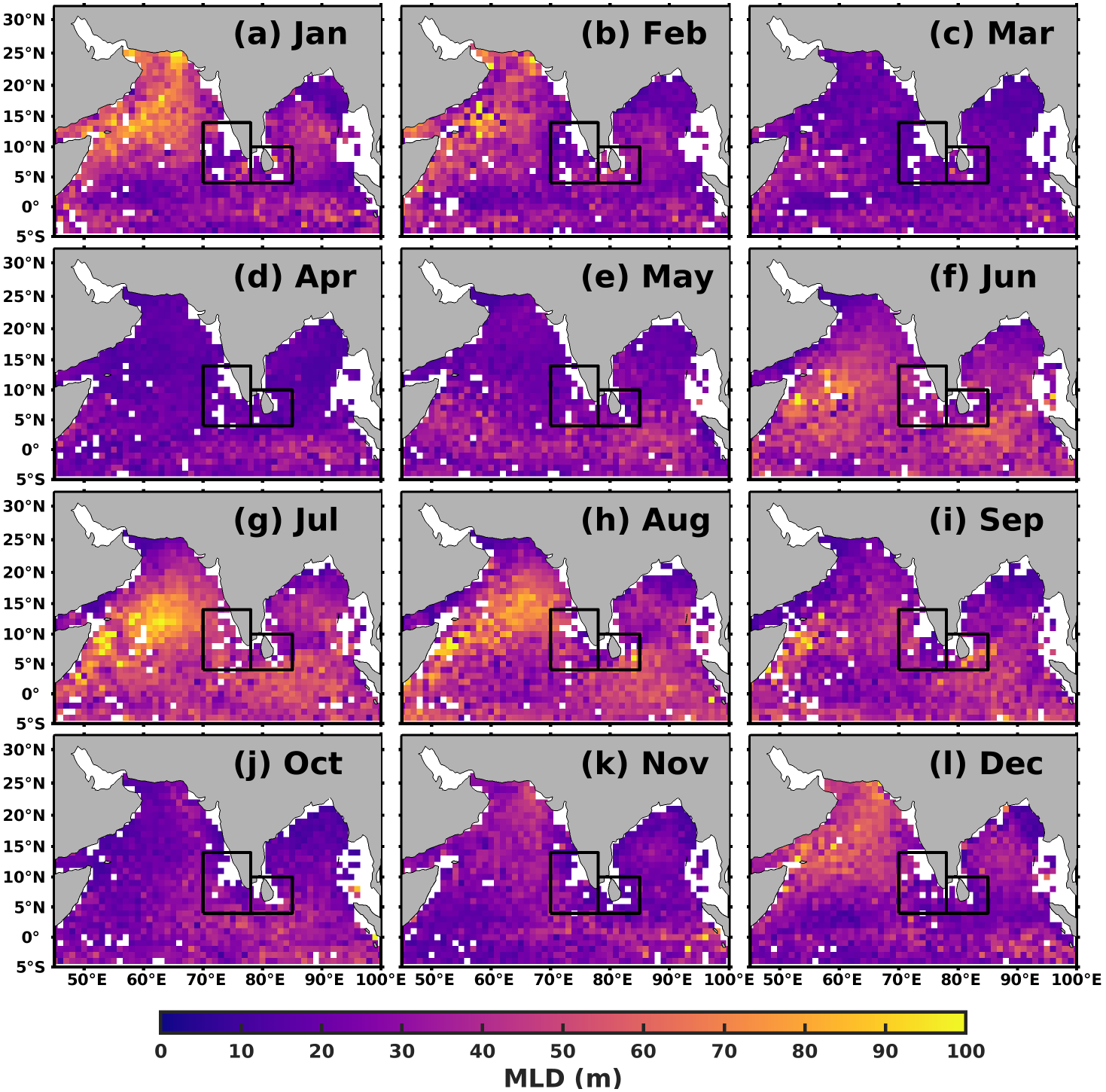}
	\caption{(a)-(l) shows the mixed-layer climatology from January to December.}
	\label{S2}
\end{figure}

\begin{figure}
	\includegraphics[width=\textwidth]{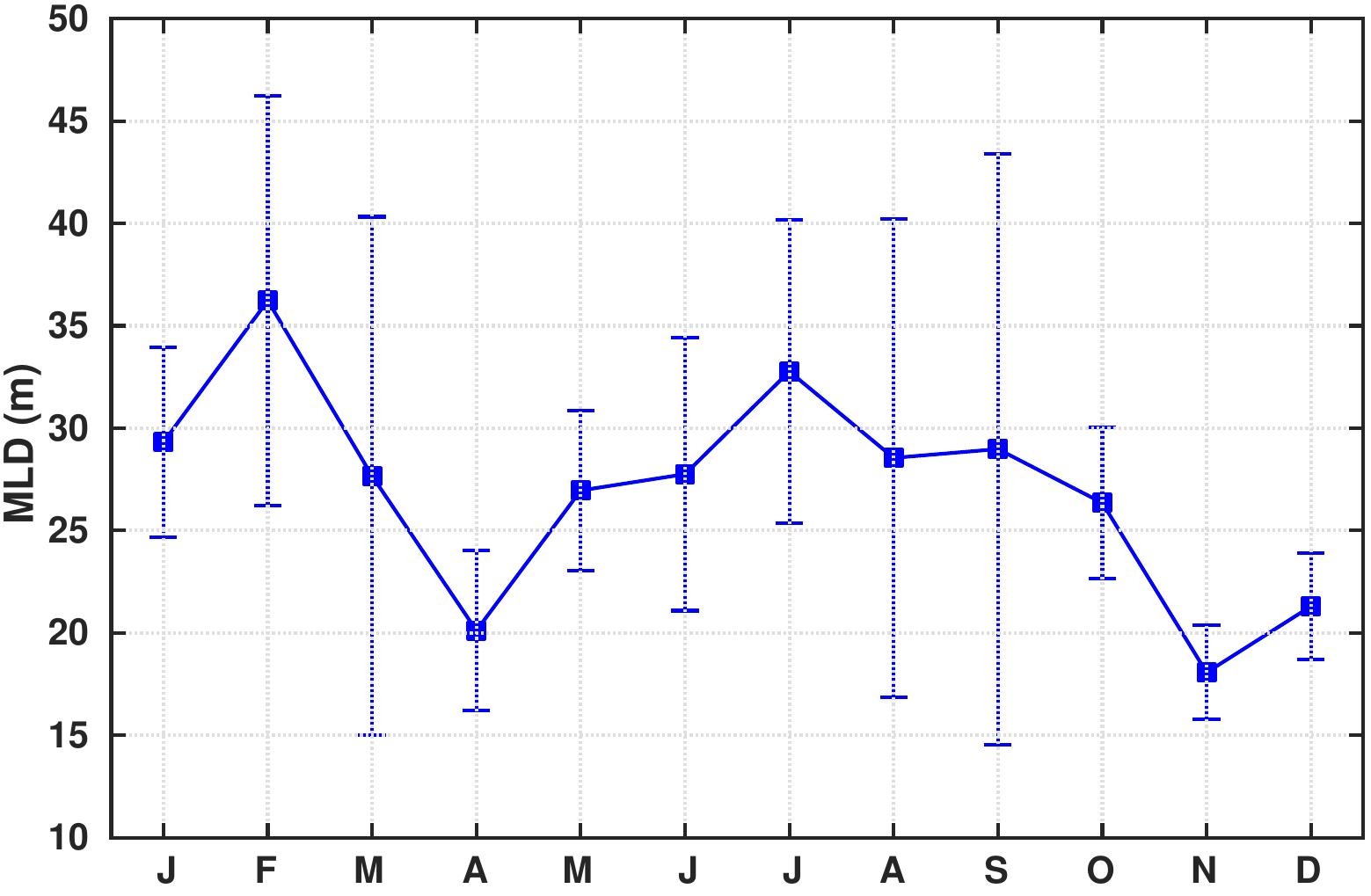}
	\caption{Monthly mixed-layer depth (MLD in meters) averaged over the channel width from 4-8$^\circ$N (blue). Error bars show the standard error of the mean (SEM) for the estimates of MLD given by $\sigma/\sqrt{N}$, where $\sigma$ and N denotes the standard deviation and number of sample points.}
	\label{S3}
\end{figure}

\begin{figure}
	\includegraphics[width=\textwidth]{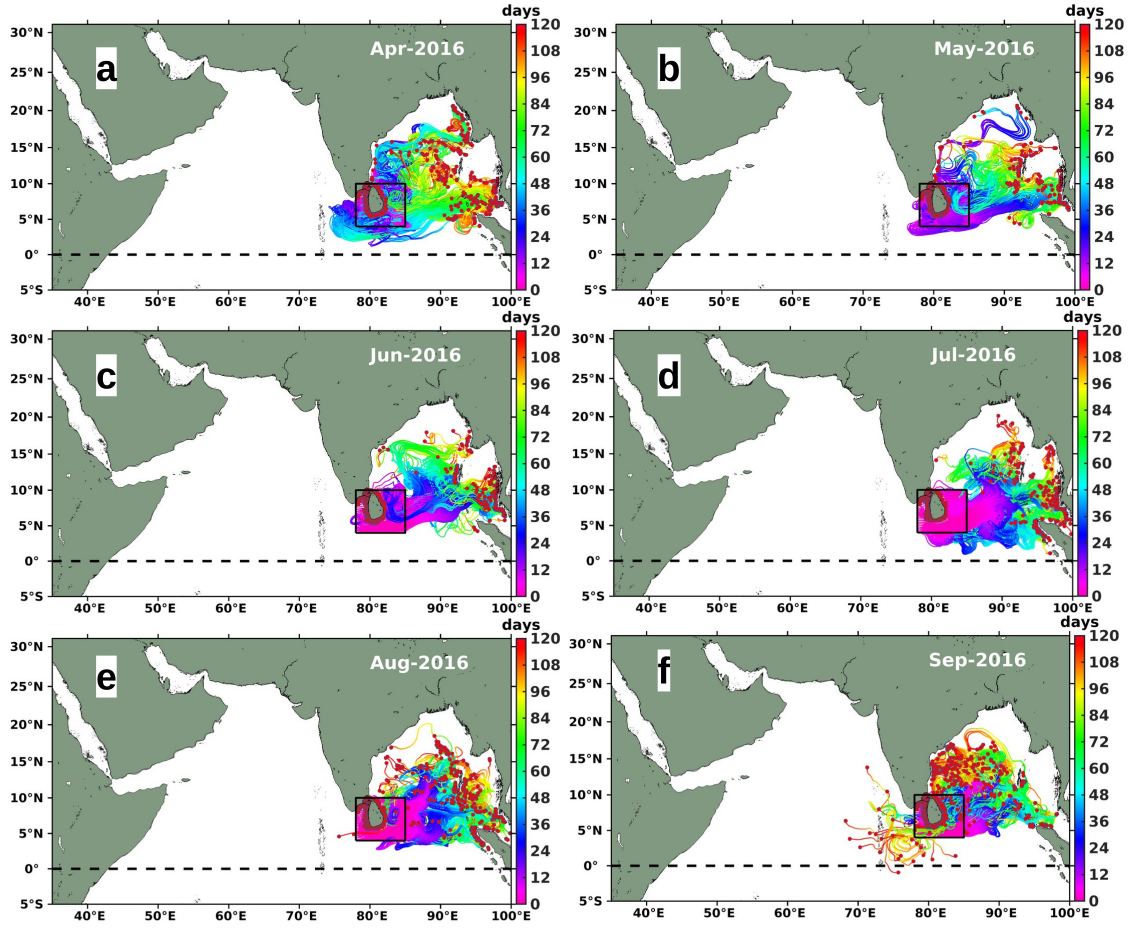}
	\caption{(a)-(f) shows the trajectory of the virtual parcels released from 1st day of April 2016 to September 2016 initialized at a resolution of $\frac{1}{4}^\circ$ in box B1 advected by surface Globcurrent velocities for 120 days (colormap showing the timespan from the initial position). The red dots are the final position of the parcel after 120 days (if the parcel ends on the coast, the preceding position is plotted).}
	\label{S4}
\end{figure}

\begin{figure}
	\includegraphics[width=\textwidth]{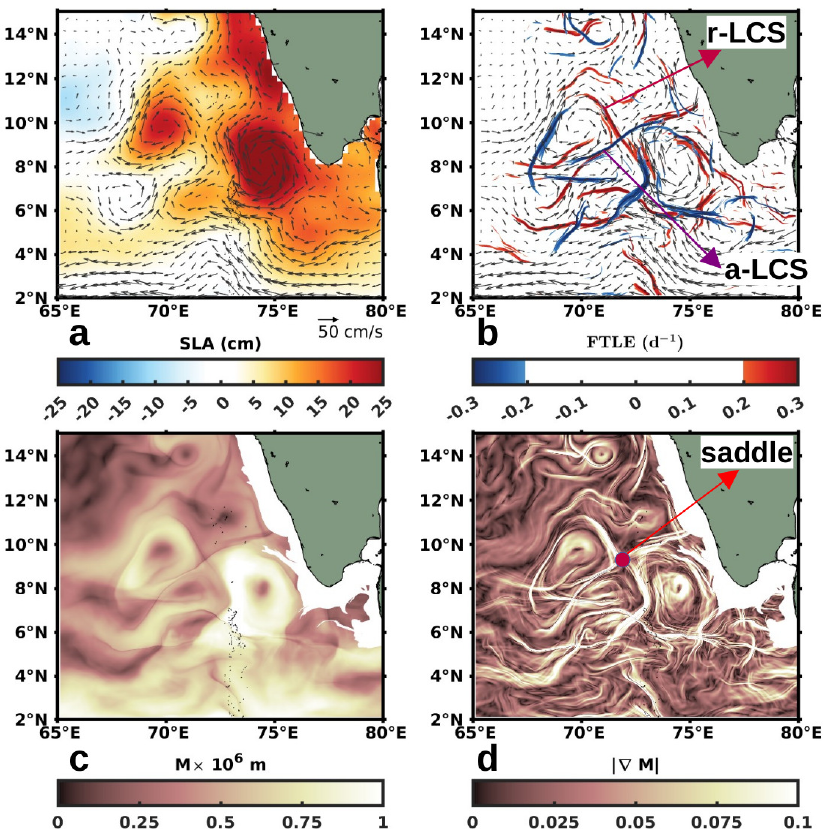}
	\caption{(a) Sea-level anomaly with geostrophic quivers; (b) b-FTLE and f-FTLE advected for two weeks with geostrophic quiver; (c) M-function and (d) gradient of M-function with a forward and backward integration time of 14 days on February 1, 2016.}
	\label{S5}
\end{figure}

	\begin{figure}
		\includegraphics[width=\textwidth]{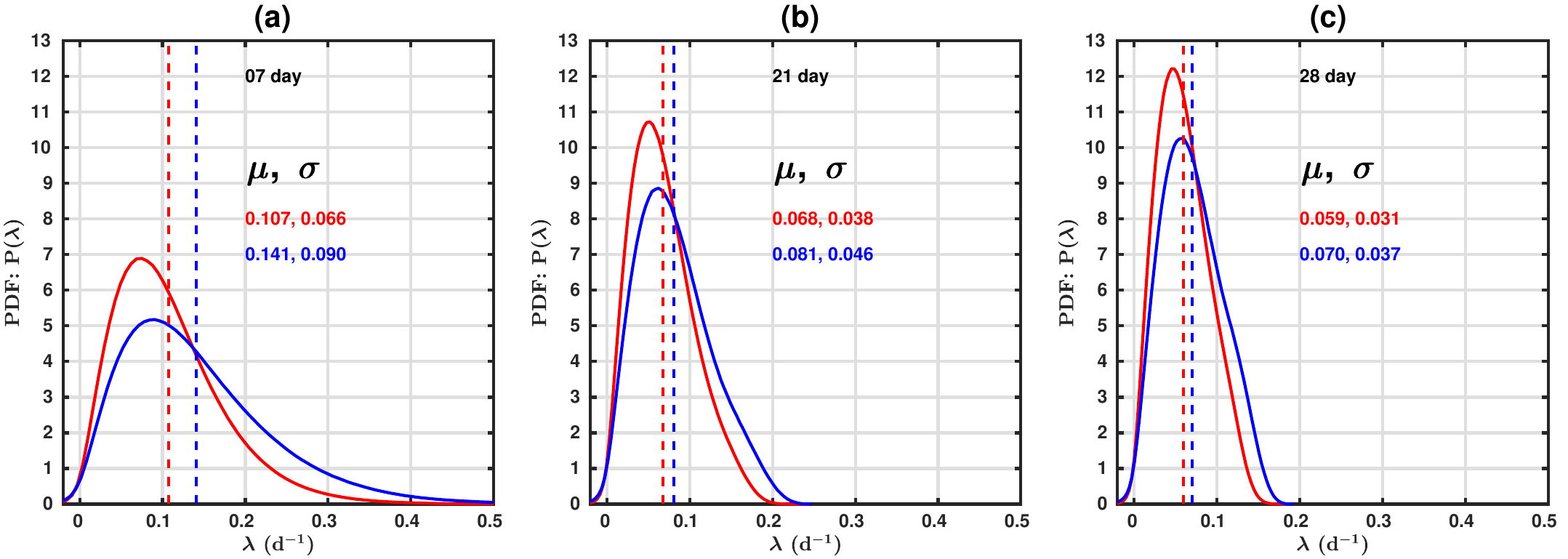}
		\caption{(a)-(c) shows the PDF of f-FTLE for the eastern (red) and western (blue) Arabian Sea (E-AS; 60.125$^\circ$E-80.125$^\circ$E, 2$^\circ$N-31.125$^\circ$N and W-AS; 45$^\circ$E-60$^\circ$E, 2$^\circ$N-31.125$^\circ$N) for the integration time of 7, 14, and 28 days with their respective mean and standard deviation (data record 2013 to 2022).}
		\label{S6}
	\end{figure}
	
	\begin{figure}[hbt]
		\centering
		\includegraphics[scale=0.95]{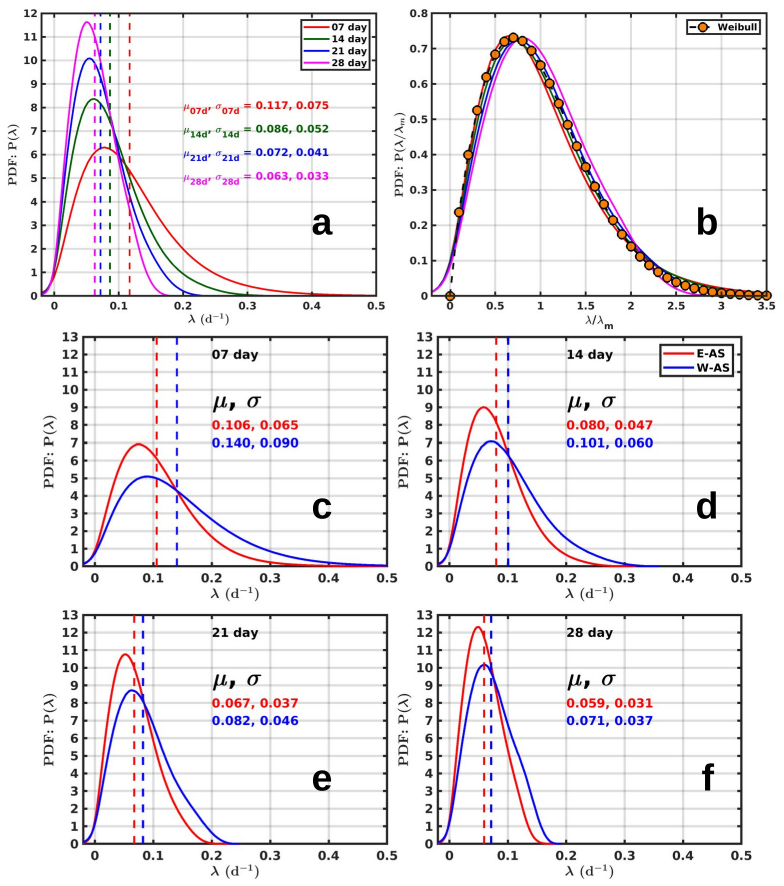}
		\caption{(a) shows Probability Density Function (PDF) of Lyapunov Exponent ($\lambda$, f-FTLE, units in day$^{-1}$) for an integration time of a week to 4 weeks estimated from the surface geostrophic current (record length beginning from 1993-2002) for the Arabian Sea; the mean ($\mu$) and the standard deviation ($\sigma$) in day$^{-1}$ are mentioned in the text. (b) shows the normalized PDF (by the mean) for all the integration times, and curves are fitted with a Weibull distribution with parameters a=1.1 and b=1.8. (c)-(f) shows the PDF of f-FTLE for the eastern and western Arabian Sea (E-AS; 60.125$^\circ$E-80.125$^\circ$E, 2$^\circ$N-31.125$^\circ$N and W-AS; 45$^\circ$E-60$^\circ$E, 2$^\circ$N-31.125$^\circ$N) for the integration time of 7, 14, 21, and 28 days with their respective mean, and standard deviation.}
		\label{S7}
	\end{figure}
	
	\begin{figure}[hbt]
		\centering
		\includegraphics[scale=0.95]{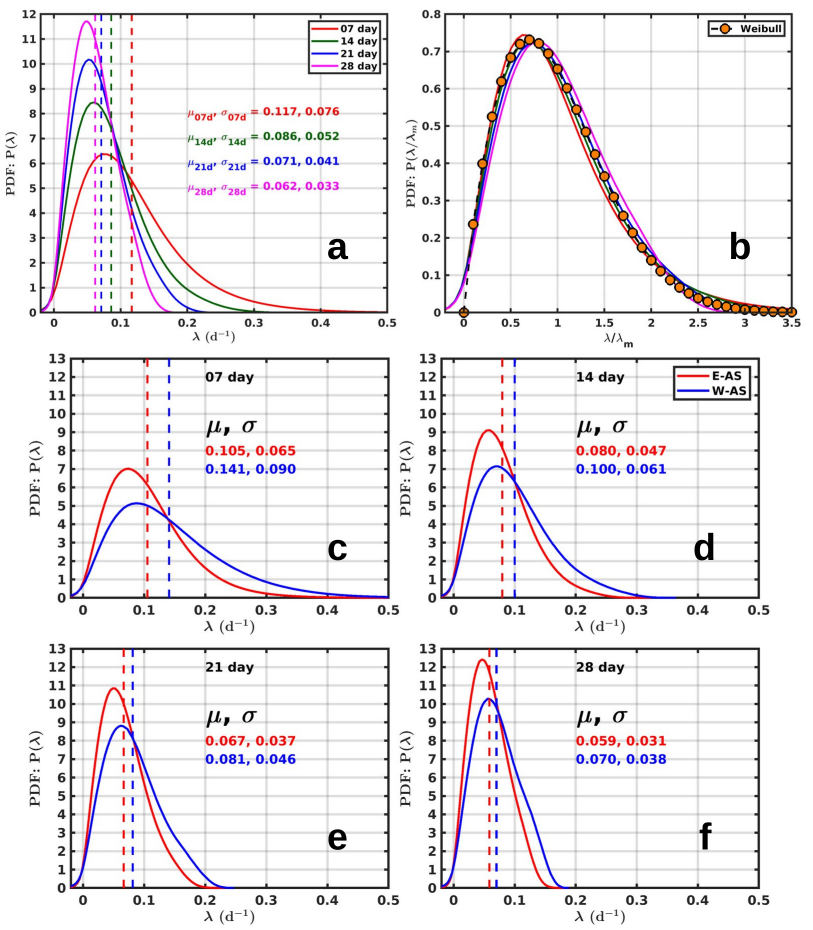}
		\caption{(a) shows Probability Density Function (PDF) of Lyapunov Exponent ($\lambda$, f-FTLE, units in day$^{-1}$) for an integration time of a week to 4 weeks estimated from the surface geostrophic current (record length beginning from 2003-2012) for the Arabian Sea; the mean ($\mu$) and the standard deviation ($\sigma$) in day$^{-1}$ are mentioned in the text. (b) shows the normalized PDF (by the mean) for all the integration times, and curves are fitted with a Weibull distribution with parameters a=1.1 and b=1.8. (c)-(f) shows the PDF of f-FTLE for the eastern and western Arabian Sea (E-AS; 60.125$^\circ$E-80.125$^\circ$E, 2$^\circ$N-31.125$^\circ$N and W-AS; 45$^\circ$E-60$^\circ$E, 2$^\circ$N-31.125$^\circ$N) for the integration time of 7, 14, 21, and 28 days with their respective mean, and standard deviation.}
		\label{S8}
	\end{figure}
	
	\begin{figure}
		\includegraphics[width=\textwidth]{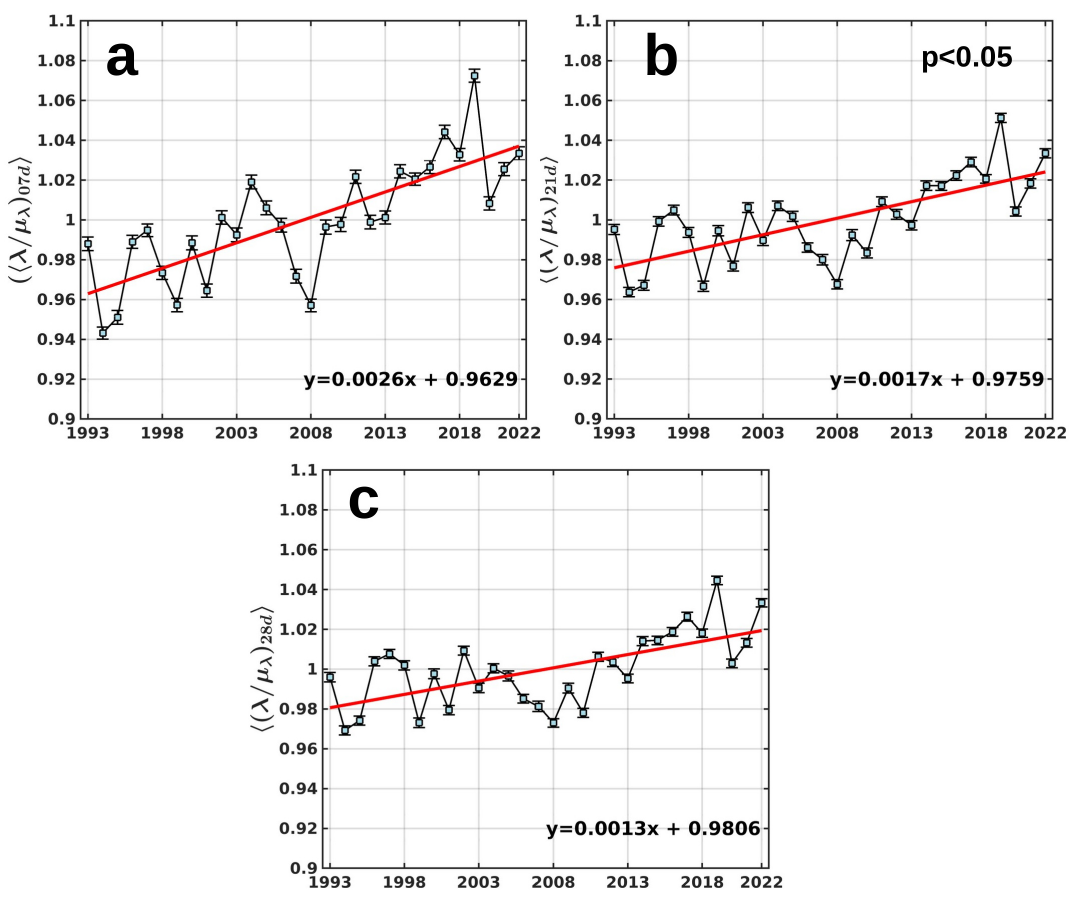}
		\caption{(a)-(f) shows the snapshots of Soil Moisture Active Passive (SMAP) sea-surface salinity (SSS) with contours from October to March (day 1) of 2015-2016.}
		\label{S9}
	\end{figure}

\end{document}